\documentstyle[preprint,aps]{revtex}

\begin{document}
\draft
\preprint{KANAZAWA 94-07,  
\ April, 1994}  
\title{
Monopoles and string tension in SU(2) QCD 
}
\author{
Hiroshi Shiba
\footnote{ E-mail address:shiba@hep.s.kanazawa-u.ac.jp}
 and Tsuneo Suzuki
\footnote{ E-mail address:suzuki@hep.s.kanazawa-u.ac.jp}
}
\address{
Department of Physics, Kanazawa University, Kanazawa 920-11, Japan
}
\maketitle

\begin{abstract}
Monopole and photon contributions to abelian Wilson loops are calculated using 
Monte-Carlo simulations of SU(2) QCD in the maximally abelian gauge. 
The string tension  is well reproduced only by  
monopole contributions, whereas photons alone are responsible for 
the Coulomb coefficient of the abelian static potential. 
\end{abstract}
       
\section{Introduction}
QCD confines color, but color confinement mechanism is not yet clarified.
A promising picture is the dual Meissner effect due to condensation 
of some color magnetic quantity
\cite{thooft1,mandel}. 
This picture is realized in the confinement phase of lattice compact
QED\cite{poly,bank,degrand}. Especially interesting 
are the following facts:\\
1) A dual transformation can be done, leading us to 
an action describing a monopole Coulomb gas
\cite{bank,peskin,frolich,smit}. Monopole condensation is shown to 
occur in the confinement phase from energy-entropy balance. \\
2) The monopole contribution alone can reproduce the full value
of the string tension\cite{stack}.

In the case of QCD, there is a difficult problem. We have to find a 
color magnetic quantity in QCD.  
The 'tHooft idea of abelian
projection of QCD\cite{thooft2} is very interesting in this respect.  
The abelian projection of QCD is to extract an abelian theory 
performing  a partial gauge-fixing  
such that the maximal abelian torus group remains unbroken. 
After the abelian projection,  $SU(3)$
QCD can be regarded as a $U(1)\times U(1)$ abelian gauge theory
with magnetic monopoles and electric charges. 't Hooft 
conjectured that the condensation 
of the abelian monopoles is the confinement mechanism in 
QCD\cite{thooft2}.

There are, however, infinite ways of extracting such an 
abelian theory out of $SU(3)$ QCD. It seems important to find a 
good gauge in which the conjecture is seen clearly to be realized.
A gauge called maximally abelian (MA) gauge has been shown to be very 
interesting \cite{kron,yotsu,suzu1}. In the MA gauge, 
there are phenomena which may be called abelian dominance
\cite{yotsu,hioki}. 
Especially the string tension in $SU(2)$ QCD is well reproduced by 
Wilson loops composed of abelian link fields alone in the MA gauge.
Moreover the monopole current 
$k_{\mu}(s)$ can be 
defined similarly as in compact QED\cite{degrand}.
It is shown in the MA gauge that the abelian monopoles
are dense and dynamical in the confinement phase, whereas they
are dilute and static in the deconfinement phase\cite{suzu1}.

Recently we have derived an effective $U(1)$ monopole action 
in the MA gauge and in $SU(2)$ QCD\cite{shiba,shiba1}. 
Entropy dominance over energy of the monopole loops, i.e.,
condensation of the monopole loops seems to occur 
in the confinement phase if extended 
monopoles\cite{ivanenko} are considered\cite{shiba,shiba1}. 
After the abelian projection in the MA gauge, infrared behaviors of 
$SU(2)$ QCD may be described by a compact-QED like $U(1)$ theory 
with the running
coupling constant instead of the bare one and with 
the monopole mass on a 
dual lattice.

If the monopoles alone are responsible for the confinement mechanism, 
the string tension which is a key quantity 
of confinement must be explained 
by monopole contributions alone. This is realized in compact QED 
\cite{stack}. 
The aim of this note is to show 
that the same thing happens also in $SU(2)$ QCD 
by means of evaluating 
monopole and photon contributions to abelian Wilson loops.
Preliminary results are reported by the present authors\cite{shiba}  and 
other group\cite{stack2}. 

\section{Abelian projection and gauge invariance}
We adopt the usual SU(2) Wilson action.
The maximally abelian  gauge is given \cite{kron} 
by performing a local gauge transformation $V(s)$ such that 

$$  R=\sum_{s,\hat\mu}{\rm Tr}\Big(\sigma_3 \widetilde{U}(s,\hat\mu)
              \sigma_3 \widetilde{U}^{\dagger}(s,\hat\mu)\Big)   $$
is maximized.
Here 
\begin{equation}
\widetilde{U}(s,\hat\mu)=V(s)U(s,\hat\mu)V^{-1}(s+\hat\mu). \label{vuv}
\end{equation}
After  the  gauge fixing is over, there still remains a $U(1)$ 
symmetry. We can extract an abelian link gauge 
variable  from the $SU(2)$ ones as follows;
\begin{equation}
   \widetilde{U}(s,\hat\mu) =
        A(s,\hat\mu)u(s,\hat\mu), \label{au}       
\end{equation}
 where $u(s,\hat\mu)$ is diagonal and  $A(s,
\hat\mu)$ has off-diagonal components. It is easy to show 
that the above fields 
$u(s,\hat\mu)$ and $A(s,\hat\mu)$ behave under the residual 
$U(1)$ transformation $d(s)$ as
 an abelian gauge field and charged matters, respectively:
\begin{eqnarray}
u(s,\hat\mu) \rightarrow u'(s,\hat\mu)=
d(s)u(s,\hat\mu)d^{\dagger}(s+\hat\mu), \label{utrans}\\ 
A(s,\hat\mu) \rightarrow A'(s,\hat\mu)=
d(s)A(s,\hat\mu)d^{\dagger}(s). \label{atrans}
\end{eqnarray}

Let us repeat that a $U(1)$ invariant 
quantity 
written in terms of the abelian link variables $u(s,\hat\mu)$
 after an abelian projection 
can  be rewritten in a  $SU(2)$ invariant (but complicated ) form 
using the original link variables $U(s,\hat\mu)$\cite{hioki}.
 The gauge function $V(s)$ which maximizes $R$ is a 
 functional of $U(s,\hat\mu)$. First study the 
transformation property  
of $V(s)$ under any $SU(2)$ transformation $W(s)$, fixing  
the $U(1)$ ambiguity of $V(s)$ in some way.
Considering that a gauge-fixed quantity does not transform, we see

\begin{equation}
  V(s) \rightarrow V^{W}(s)= d(s)V(s)W^{-1}(s) 
\label{vw},
\end{equation}
where $d(s)\  (\in U(1) )$ 
ensures the form invariance of $V^{W}(s)$ and $V(s)$ and 
is determined uniquely by $V(s)$ and $W(s)$.
Using the definition (\ref{vuv}) and the transformation property (\ref{vw}),
 we get 
\begin{equation}
 (\widetilde{U}(s,\hat\mu))^{W} = d(s)\widetilde{U}(s,\hat\mu)
d^{\dagger}(s+\hat\mu) \label{uw}
\end{equation}
for any $SU(2)$ transformation $W(s)$.  Hence all $U(1)$ invariant quantities 
composed of $\widetilde{U}(s,\hat\mu)$ (and  $u(s,\hat\mu)$)
 are automatically $SU(2)$ invariant after the abelian projection.

As an example, 
consider a $U(1)$-invariant $1 \times 1$ plaquette
variable $u_P(s,\mu,\nu)$ composed of $u(s,\hat\mu)$ alone.
Using (\ref{vuv}) and (\ref{au}), we get
\begin{eqnarray}
u_P(s,\mu,\nu) &=& \frac{1}{2} {\rm Tr}(u(s,\hat\mu)u(s+\hat\mu,\hat\nu)
u^{\dagger}(s+\hat\nu,\hat\mu)u^{\dagger}(s,\hat\nu))\\
&=&  \frac{1}{2} {\rm Tr}(U'(s,\hat\mu)U'(s+\hat\mu,\hat\nu)
U'^{\dagger}(s+\hat\nu,\hat\mu)U'^{\dagger}(s,\hat\nu)),
\end{eqnarray}
where $U'(s,\hat\mu)$ is a modified link field defined by 
 $U'(s,\hat\mu)=B(s,\hat\mu)U(s,\hat\mu)$
and 
$B(s,\hat\mu)= V^{\dagger}(s)A^{\dagger}(s,\hat\mu)V(s)$.
Under an arbitrary $SU(2)$ transformation $W(s)$, 
we see from (\ref{atrans}), (\ref{vw}) and (\ref{uw})
\begin{eqnarray}
B(s,\hat\mu) \rightarrow W(s)B(s,\hat\mu)W^{\dagger}(s),\\
U'(s,\hat\mu) \rightarrow W(s)U'(s,\hat\mu)W^{\dagger}(s+\hat\mu).
\end{eqnarray}
Namely, $U'(s,\hat\mu)$ transforms like the corresponding original link 
field $U(s,\hat\mu)$.
The $SU(2)$ invariance of 
 $u_P(s,\mu,\nu)$ is seen also from this property. Moreover, it is  seen
how different the abelian plaquette variable is from the full one. 
 $u_P(s,\mu,\nu)$ in a different abelian projection corresponds to 
a differently modified  full plaquette variable composed of 
 $U'(s,\hat\mu)$.

\section{Monopole and photon contributions to abelian Wilson loops}
We show an abelian Wilson loop operator 
written in terms of $u(s,\hat\mu)$ alone after the abelian projection is 
given  by a product of monopole and photon contributions. Here
we take into account only a simple Wilson loop, say, of size
$I \times J$. Then such an 
abelian Wilson loop operator is expressed as 
\begin{eqnarray}
W = \exp\{i\sum J_{\mu}(s)\theta_{\mu}(s)\}, 
\end{eqnarray}
where  $J_{\mu}(s)$ 
is an external current taking $\pm 1$ along the Wilson loop
 and $\theta_{\mu}(s)$ is an angle variable defined from  $u(s,\hat\mu)$ 
as follows:
\begin{eqnarray}
u(s,\hat\mu)= \left( \begin{array}{cc}
e^{i\theta_{\mu}(s)} & 0 \\
0 & e^{-i\theta_{\mu}(s)} 
\end{array} \right).
\end{eqnarray}
 Since $J_{\mu}(s)$
is conserved, it is rewritten for such a simple Wilson loop 
in terms of an antisymmetric  
variable $M_{\mu\nu}(s)$ 
as $J_{\nu}(s)=\partial_{\mu}'M_{\mu\nu}(s)$, 
where $\partial'$ is a backward 
derivative on a lattice. 
$M_{\mu\nu}(s)$ takes $\pm 1$ on a surface with the 
Wilson loop boundary.
Although we can choose any surface of such a type, 
we adopt a minimal 
surface here.  
We get 
\begin{equation}
W  =  \exp \{-\frac{i}{2}\sum M_{\mu\nu}(s)f_{\mu\nu}(s)\}
\label{W},
\end{equation}
where $f_{\mu\nu}(s)= \partial_{\mu}\theta_{\nu}(s) - 
\partial_{\nu}\theta_{\mu}(s)$ and $\partial_{\mu}$ is a 
forward derivative on a lattice.
The gauge plaquette variable can be decomposed into $f_{\mu\nu}(s) = 
\bar{f}_{\mu\nu}(s)+ 2\pi n_{\mu\nu}(s)\ $  where 
$\bar{f}_{\mu\nu}(s) \in [-\pi, \pi] $ corresponds to a 
field strength and $n_{\mu\nu}(s)$ 
is an integer-valued plaquette variable
denoting the Dirac string 
\footnote{ This condition 
$\bar{f}_{\mu\nu}(s) \in [-\pi, \pi] $
is applicable only in the case of
defining a smallest $1^3$ monopole.}.  
Since $M_{\mu\nu}(s)$ and $n_{\mu\nu}(s)$ are integers, the latter does not 
contribute to Eq.\ (\ref{W}). Hence $f_{\mu\nu}(s)$ in 
Eq.\ (\ref{W}) is replaced by
$\bar{f}_{\mu\nu}(s)$. 
Using a decomposition rule 
\begin{eqnarray*}
M_{\mu\nu}(s) = -\sum D(s-s')[\partial'_{\alpha}(\partial_{\mu}
M_{\alpha\nu}-\partial_{\nu}M_{\alpha\mu})(s')\\
+ \frac{1}{2}\epsilon_{\alpha\beta\mu\nu}
\epsilon_{\alpha '\beta\rho\sigma}
\partial'_{\alpha}\partial_{\alpha '}M_{\rho\sigma}(s')],
\end{eqnarray*}
we get 
\begin{eqnarray}
W\    & = & W_{1} \cdot W_{2} \label{w12}\\
W_{1} & = & \exp\{-i\sum \partial'_{\mu}\bar{f}_{\mu\nu}(s)
D(s-s')J_{\nu}(s')\} \nonumber \\
W_{2} & = & \exp\{2\pi i\sum k_{\beta}(s)D(s-s')\frac{1}{2}
\epsilon_{\alpha\beta\rho\sigma}\partial_{\alpha}M_{\rho\sigma}(s')\}, 
\nonumber
\end{eqnarray}
where a monopole current $k_{\mu}(s)$ is defined as $k_{\mu}(s)= 
(1/4\pi)\epsilon_{\mu\alpha\beta\gamma}\partial_{\alpha}
\bar{f}_{\beta\gamma}(s)$ following DeGrand-Toussaint\cite{degrand}.
$D(s)$ is the lattice Coulomb propagator. 
Since $\bar{f}_{\mu\nu}(s)$ corresponds to the field strength of the 
photon field, 
$W_{1} (W_{2})$ 
is the photon (the monopole) contribution to the Wilson loop.
To study the features of both contributions, 
we evaluate the expectation 
values $\langle W_1 \rangle$
and $\langle W_2 \rangle$ separately and compare them with those 
of $W$. 

\section{Simulations and extended monopoles }
The Monte-Carlo simulations were done on $24^4$ lattice from $\beta =
2.4$ to $\beta =2.7$. All measurements were done every 30 sweeps after
a thermalization of 1500 sweeps. We took 50 
configurations totally for
measurements. The gauge-fixing criterion 
is the same  as done in Ref.\ \cite{ohno}. 
Using gauge-fixed configurations, we evaluated monopole 
currents and obtained the ensemble of monopole currents. 

As shown in the previous note\cite{shiba,shiba1}, 
type-2 extended monopole 
loops with $b > b_c \sim 5.2\times 10^{-3}(\Lambda_L)^{-1}$ 
seem to condense, where $b=na(\beta)$ for $n^3$ extended monopoles 
and $a(\beta)$ is the lattice constant. 
So we measure $2^3$ ($3^3$) extended 
monopoles with $b=2a(\beta) (b=3a(\beta))$ of the type-2\cite{ivanenko}
as well as usual smallest ones. The extended monopole of the type-2 
is defined on an extended cube as the sum of the smallest ones included 
in the cube as shown in Fig.\ \ref{cube}.
Note that the definition of the type-2 extended monopoles 
corresponds to making a block
spin transformation of the monopole currents with the scale factor $n$.
Hence the effective lattice volume is reduced.
We call the effective lattice as a renormalized lattice.
We evaluate the averages of $W$ using abelian 
link variables  (called abelian) on the original lattice, 
of $W_1$ (photon part),  and $W_2$ (monopole 
part) on each renormalized lattice, separately. 

\section{Results}
The results are shown in the following:
\begin{enumerate}
\item
The monopole contributions to Wilson loops are 
obtained with  relatively 
small errors. Surprisingly enough, 
the Creutz ratios $\chi (I,J)$ of the monopole contributions 
having small errors  
are almost independent of the loop size for example as shown 
in Fig.\ \ref{chi}.
This means that the monopole contributions are composed  only 
of an area, a perimeter, and a constant terms without a Coulomb term. 
\item
Assuming the static potential is given 
by a linear + Coulomb + constant
terms, we try to determine the potential using the 
least square fit. There are various ways, but we adopt a method similar to
that\cite{barkai} using the Creutz ratios.
The assumption of the form of the static potential leads us to the Creutz ratio
\begin{equation}
\chi (I,J)  =  \chi_0 - \chi_1 (\frac{1}{I(I-1)}+\frac{1}{J(J-1)})
+\chi_2 (\frac{1}{I(I-1)J(J-1)}),
\label{ch}
\end{equation}
where $\chi_0$ is the string tension and $\chi_1$ corresponds 
to the Coulomb coefficient of the static potential. Using 
the fitted values of $\chi_0$ and $\chi_1$, we can reproduce each static 
potential.

We plot their data in Fig.\ \ref{pot25} 
(at $\beta =2.5$) and in 
Fig.\ \ref{pot26} (at $\beta =2.6$). 
We find the monopole contributions are responsible for
the linear-rising behaviors. 
When the $2^3$ extended monopoles are used, 
we obtain almost the same results.
The photon part contributes only to the 
short-ranged region. 
There seems to exist a small discrepancy between the 
abelian and the monopole + photon parts for $R/a =12$, 
but  finite-size effects are expected there. Moreover, the assumption
of the form of the static potential looks too simple. 
Similar data are obtained for $\beta =2.4$ and $2.7$.
\item
This is seen more clearly from the data of the string tensions 
which are determined from the static potentials. They are shown in
Fig.\ \ref{sigma}. 
The full and the abelian string tensions have large errors but they 
are seen to be well reproduced by the 
monopoles alone for $\beta \le
2.7$ and the photon part has almost vanishing string tensions.
\item
The string tensions from the monopoles and the photons 
of various sizes are plotted in Fig.\ \ref{sigma2}.
It is interesting to see that they are almost independent 
of the extendedness 
contrary to our preliminary data on a smaller lattice\cite{ilyar}, 
although the monopole actions determined 
in \cite{shiba,shiba1} depend on the extendedness. Note that the extended 
monopoles of the type-2 are composed of the sum of the smallest 
monopoles. Hence even when the extended monopoles alone look to condense, 
the smallest monopoles also have some information of the condensation.
\item
We have derived also  Coulomb coefficients from the static potentials 
as shown in Fig.\ \ref{coulomb}. 
The monopole part has almost vanishing Coulomb 
coefficients which is in agreement with the constant behaviors of 
the Creutz rations of the monopole part as 
shown above. The $1^3$ photon part has 
large coefficients and they reproduce well the coefficients of 
the abelian static potentials.
\item
We  also measured the Coulomb coefficients of the photon parts on each 
renormalized lattice. The values depend on the extendedness. But
the following fact is interesting.
The photon parts are evaluated on a renormalized 
lattice with $b=na(\beta)$.
Hence they have different values of $b$ for different extendedness.
The Coulomb coefficients of the photon parts are well reproduced 
by the $SU(2)$ running coupling constants $g(b)$ with $b=na(\beta)$, 
i.e., $-g(b)^2/16\pi$, where 
\begin{eqnarray}
g^{-2}(b)= \frac{11}{24\pi^2}\ln(\frac{1}{b^2\Lambda^2}) 
+ \frac{17}{44\pi^2}
\ln\ln(\frac{1}{b^2\Lambda^2}).
\end{eqnarray}
The scale parameter $\Lambda$ determined is 
$\Lambda \sim 48\Lambda_L $ 
which is 
quite near to the value $\Lambda \sim 42\Lambda_L$ 
fixed from the monopole action\cite{shiba}.
\end{enumerate}

In conclusion,  our analyses done here strongly suggest 
that abelian monopoles are responsible for 
confinement in $SU(2)$ QCD 
and condensation of the monopoles is the confinement mechanism 
if the abelian projection is done in the MA gauge. 
To extend our method to a finite-temperature system and also to
$SU(3)$ with or without dynamical quarks is  
very interesting. 
These studies are also in progress.

We wish to acknowledge Yoshimi Matsubara 
for useful discussions.
This work is financially supported by JSPS 
Grant-in Aid for Scientific  
Research (c)(No.04640289).

\begin{figure}
\caption{
An extended cube on which an $2^3$ extended monopole is defined as the sum of 
eight (=$2^3$) smallest monopoles. 
}
\label{cube}
\end{figure}

\begin{figure}
\caption{Creutz ratios $\chi (I,J)$ from abelian and $2^3$ monopole 
Wilson loops at $\beta =2.6$ versus $I \times J$. 
The monopole Creutz ratio values are devided by 
$4$, being adjusted to those 
in unit $a(\beta)$. }
\label{chi}
\end{figure}

\begin{figure}
\caption{
Static potentials $aV(R)$ versus $R/a$ at $\beta =2.5$.
The values are shifted by a constant.}
\label{pot25}
\end{figure}

\begin{figure}
\caption{
Static potentials $aV(R)$ versus $R/a$ at $\beta =2.6$.
The values are shifted by a constant.}
\label{pot26}
\end{figure}

\begin{figure}
\caption{
String tensions at $\beta =2.4, 2.5, 2.6$, and $2.7$.}
\label{sigma}
\end{figure}

\begin{figure}
\caption{
String tensions on each renormalized lattice.} 
\label{sigma2}
\end{figure}

\begin{figure}
\caption{
Coulomb coefficients.}
\label{coulomb}
\end{figure}

\begin{figure}
\caption{
Coulomb coefficients on each renormalized lattice.}
\label{coulomb2}
\end{figure}


\begin{references}
\bibitem{thooft1} G.'tHooft, in {\it High Energy Physics}, edited by 
A.Zichichi (Editorice Compositori, Bologna, 1975).
\bibitem{mandel} S. Mandelstam, Phys. Rep. {\bf 23C}, (1976) 245.
\bibitem{poly} A.M. Polyakov, Phys. Lett. {\bf B59}, (1975) 82.
\bibitem{bank} T.Banks $et al.$, Nucl. Phys. {\bf B129}, (1977) 493.
\bibitem{degrand} T.A. DeGrand and D. Toussaint, 
Phys. Rev. {\bf D22}, (1980) 2478.
\bibitem{peskin} M.E. Peshkin, Ann. Phys. {\bf 113}, (1978) 122.
\bibitem{frolich} J. Fr\"{o}lich and P.A. Marchetti, Euro. Phys. Lett.
{\bf 2}, (1986) 933.
\bibitem{smit} J. Smit and A.J. van der Sijs, Nucl. Phys. 
{\bf B355}, (1991) 603; Nucl. Phys. B(Proc. Suppl.) 
{\bf 20}, (1991) 221.
\bibitem{stack} J.D. Stack and R.J. Wensley, 
Nucl. Phys. {\bf B371}, (1992) 597.
\bibitem{thooft2} G. 'tHooft, Nucl. Phys. {\bf B190}, (1981) 455.
\bibitem{kron} A.S. Kronfeld et al., Phys. Lett. {\bf B198}, 
(1987) 516;
A.S. Kronfeld et al., Nucl.Phys. {\bf B293}, (1987) 461.
\bibitem{yotsu} T. Suzuki and I. Yotsuyanagi, 
Phys. Rev. {\bf D42}, (1990) 4257.
\bibitem{suzu1} T. Suzuki, Nucl. Phys. B(Proc. Suppl.) 
{\bf 30}, (1993) 176 and see also references therein.  
\bibitem{hioki} S. Hioki et al., Phys. Lett. {\bf B272}, (1991) 326.
\bibitem{shiba} H.Shiba and T.Suzuki, 
Kanazawa University Report KANAZAWA 93-09 and 93-10, 1993 (unpublished); 
Lattice 93 report published in Nucl. Phys. B(Proc. Suppl.).
\bibitem{shiba1} H.Shiba and T.Suzuki, 
in preparation.
\bibitem{ivanenko} T.L. Ivanenko et al., Phys. Lett. {\bf B252}, 
(1990) 631.
\bibitem{stack2} J.D. Stack and R.J. Wensley, 
Lattice 93 report published in Nucl. Phys. B(Proc. Suppl.).
\bibitem{ohno} S.Hioki et al.,Phys.Lett. {\bf B271}, (1991) 201.  
\bibitem{barkai} D.Barkai et al., Phys. Rev. {\bf D30}, (1984) 1293.
\bibitem{ilyar} S.Ohno et al., Nucl. Phys. B(Proc. Suppl.) 
{\bf 30}, (1993) 561.  
\end{references}
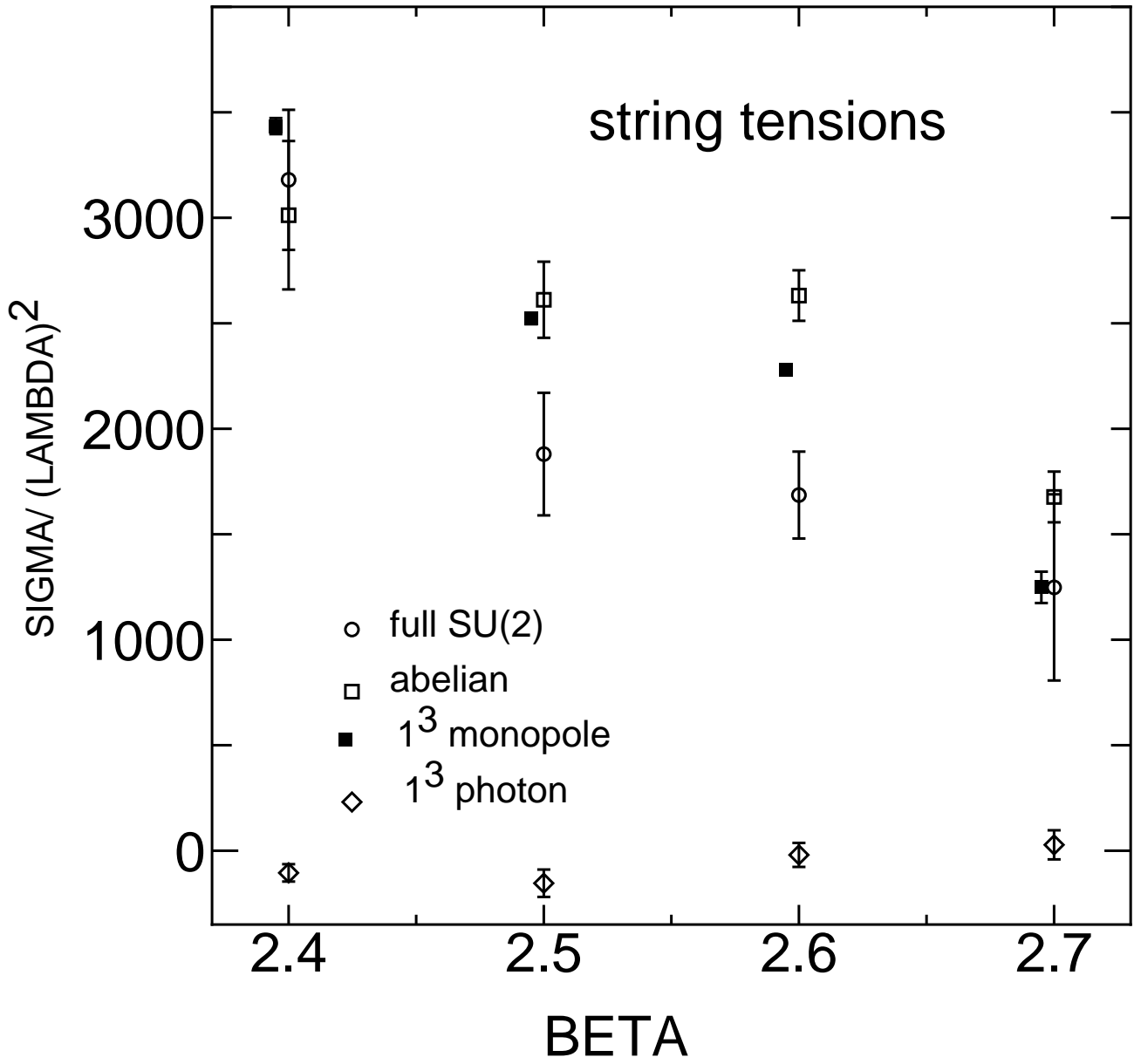
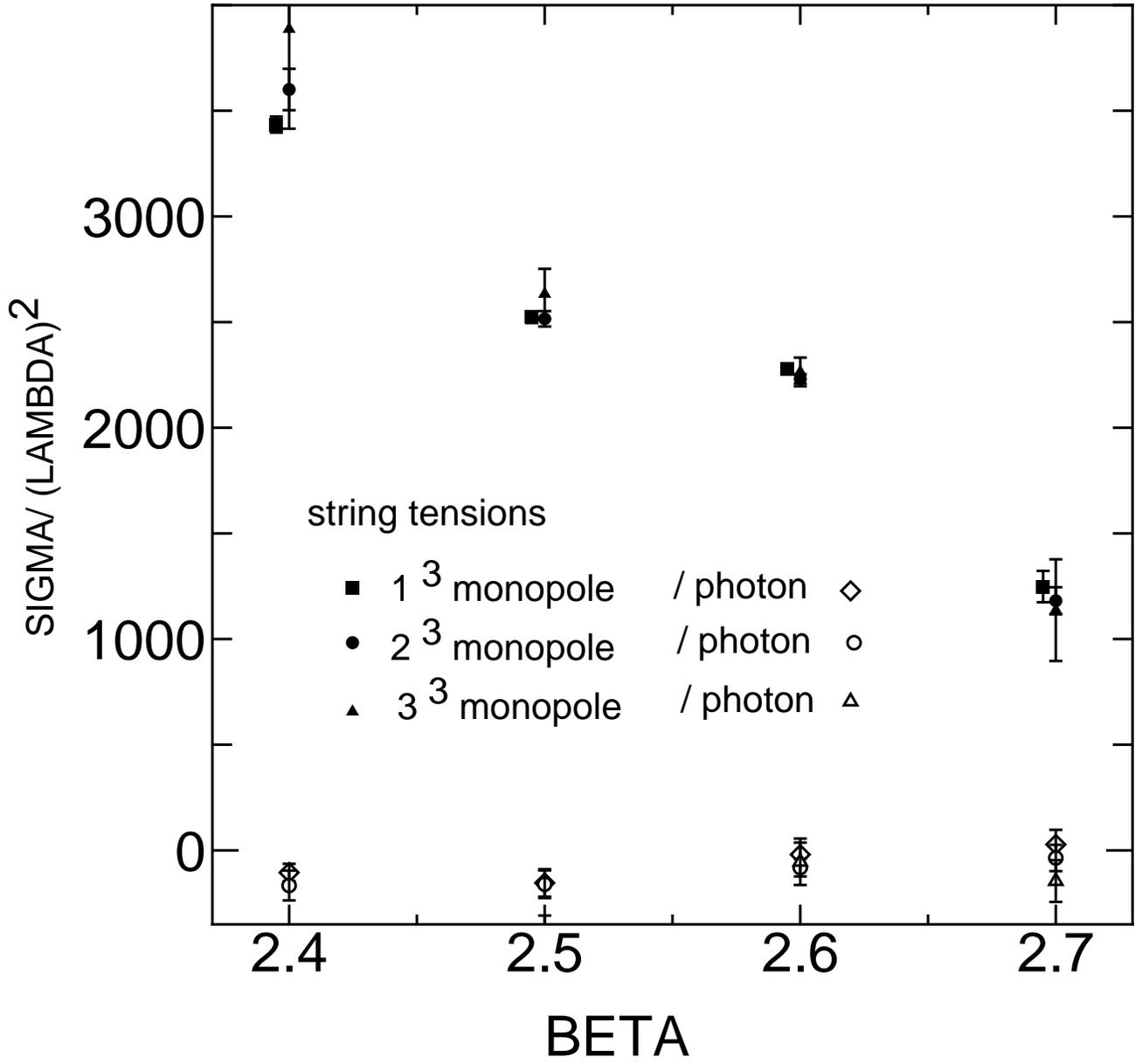
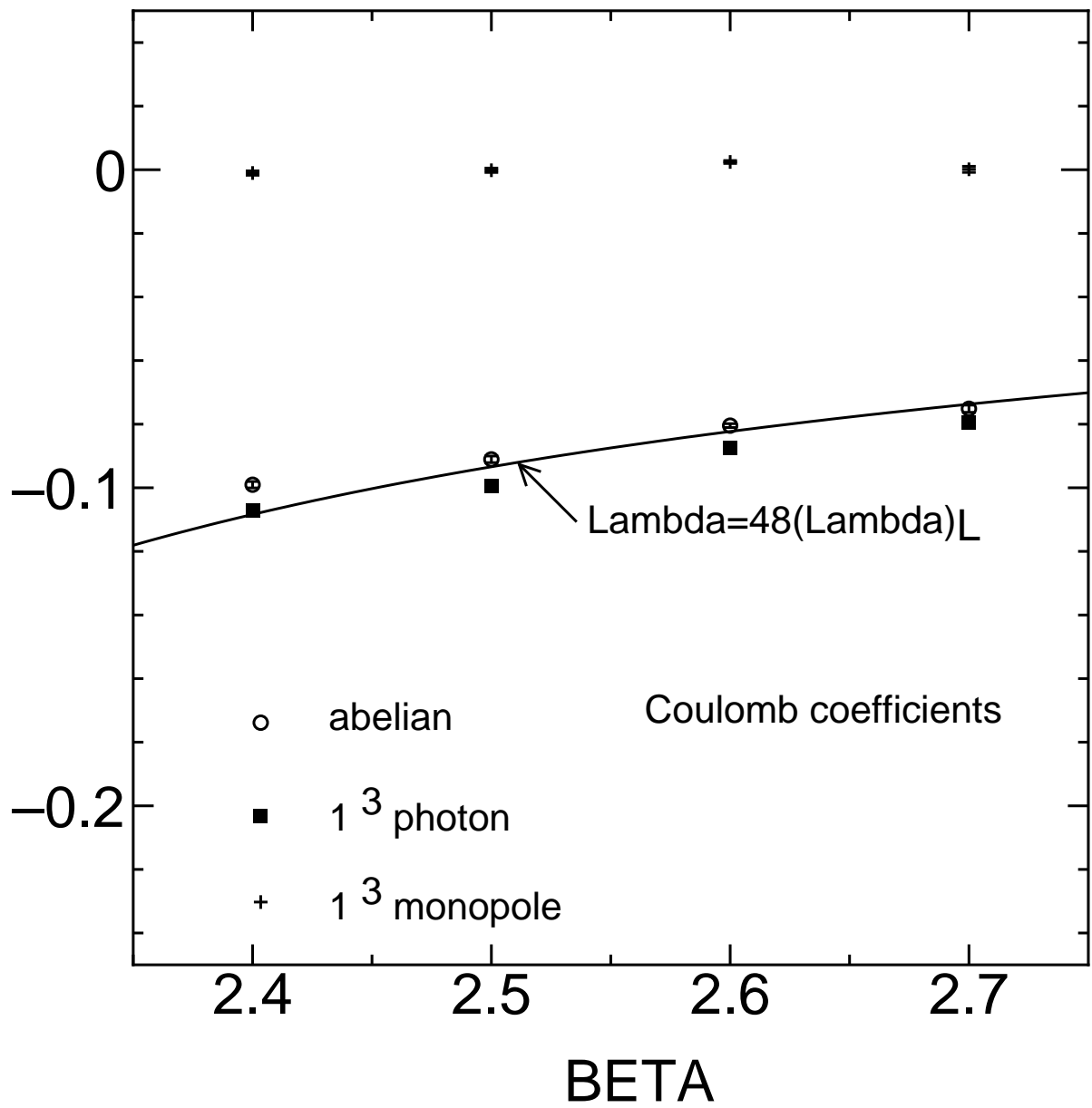
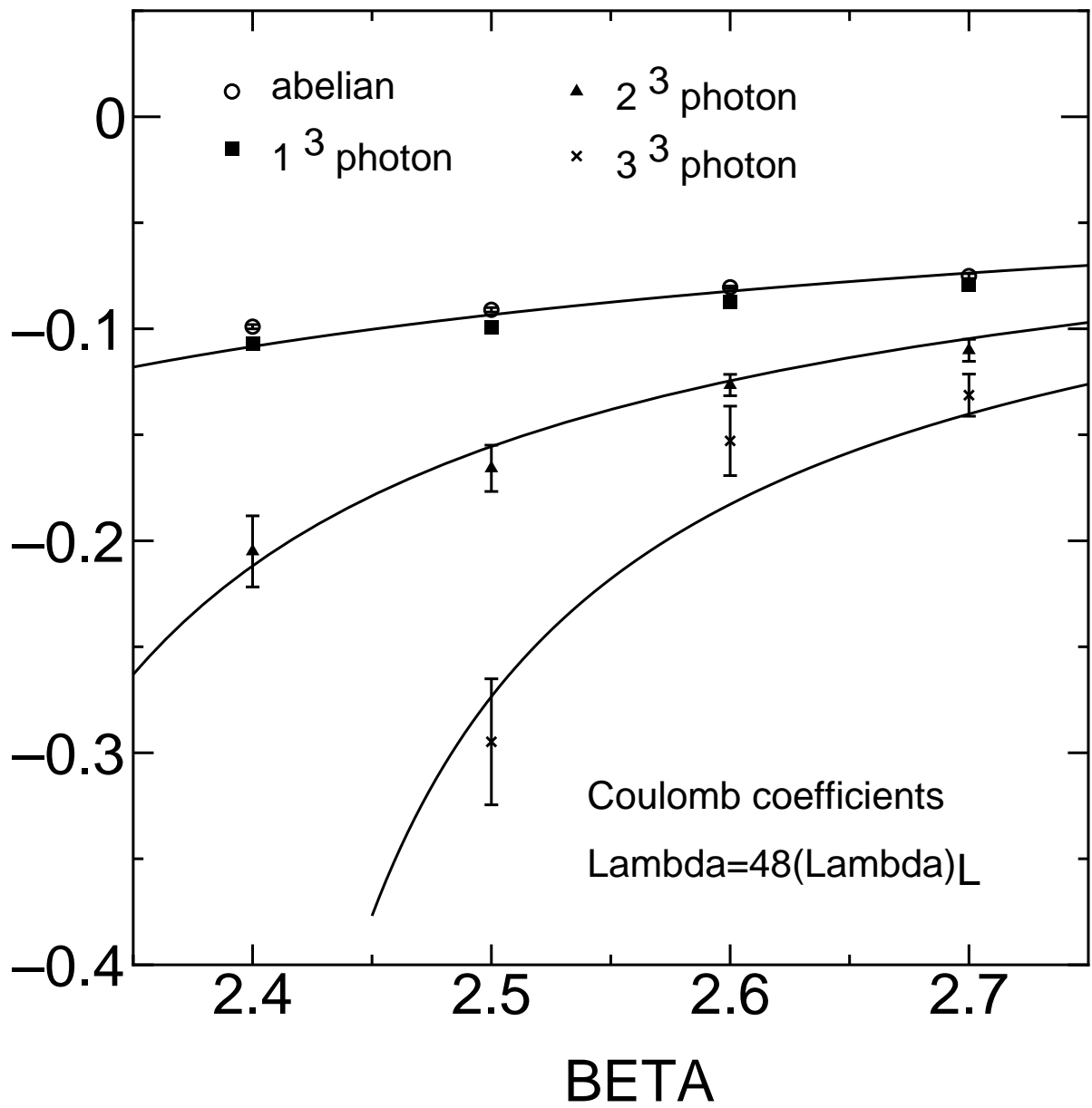
\end{document}